\documentclass{article}
\usepackage{spconf,amsmath,graphicx}
\usepackage{tabularx}
\usepackage{ragged2e}
\usepackage{lipsum}
\usepackage{bbding,amssymb}
\usepackage[misc]{ifsym}
\usepackage{pifont}

\title{Neural2Speech: A Transfer Learning Framework for Neural-Driven Speech Reconstruction}
\name{Jiawei Li$^{1}$, Chunxu Guo$^{1}$, Li Fu$^{2}$, Lu Fan$^{2}$, Edward F. Chang$^{3}$, Yuanning Li$^{1 \text{ \Letter}}$
\thanks{This work is supported by NSFC General Program 32371154 and Shanghai Pujiang Program 22PJ1410500. \Letter \ Correspondence to: Yuanning Li (liyn2@shanghaitech.edu.cn).}
}
\address{$^{1}$School of Biomedical Engineering, ShanghaiTech University, Shanghai, China.\\
$^{2}$JD AI Research, Beijing, China.\\
$^{3}$Department of Neurological Surgery, University of California, San Francisco, CA, USA.
}

\begin{document}
\ninept
\maketitle
\begin{abstract}
Reconstructing natural speech from neural activity is vital for enabling direct communication via brain-computer interfaces. Previous efforts have explored the conversion of neural recordings into speech using complex deep neural network (DNN) models trained on extensive neural recording data, which is resource-intensive under regular clinical constraints. However, achieving satisfactory performance in reconstructing speech from limited-scale neural recordings has been challenging, mainly due to the complexity of speech representations and the neural data constraints. To overcome these challenges, we propose a novel transfer learning framework for neural-driven speech reconstruction, called Neural2Speech, which consists of two distinct training phases. First, a speech autoencoder is pre-trained on readily available speech corpora to decode speech waveforms from the encoded speech representations. Second, a lightweight adaptor is trained on the small-scale neural recordings to align the neural activity and the speech representation for decoding. Remarkably, our proposed Neural2Speech demonstrates the feasibility of neural-driven speech reconstruction even with only 20 minutes of intracranial data, which significantly outperforms existing baseline methods in terms of speech fidelity and intelligibility.
\end{abstract}
\section{Introduction}
\label{sec:intro}
\begin{sloppypar}

Neural-driven speech reconstruction represents an innovative inverse mapping technique designed to approximate natural speech patterns by decoding the neural activity elicited in the brain \cite{pasley2012}. This pioneering approach has demonstrated significant potential within the realm of brain-computer interfaces, particularly in assisting individuals with paralysis and facilitating direct communication capabilities \cite{wolpaw2002brain, chang2020toward}. Conventionally, electrodes are surgically implanted as sensors to capture the brain's electrical activity directly from the cerebral cortex. Subsequently, these neural recordings are decoded to regenerate coherent speech patterns. Recent advancements in neural recording technology, characterized by high spatiotemporal resolution and a superior signal-to-noise ratio, such as electrocorticography (ECoG), have opened up exciting prospects for the field of speech reconstruction from neural activity \cite{chang2015towards}.

Speech decoding and reconstruction from intracranial recorded neural activity is a challenging sequence-to-sequence learning task and has been investigated in a number of studies in the literature. The performance of speech reconstruction is constrained by the trade-off between the amount of high-quality intracranial neural data and the complexity of the decoding algorithm. Collecting extensive intracranial neural data from human subjects engaged in speech-related tasks is often challenging and resource-intensive due to clinical constraints. On the other hand, the mapping between brain activity and the acoustic speech signal is highly nonlinear \cite{yi2019encoding}. Achieving high-quality speech reconstructions would require complex nonlinear models, such as deep neural networks. A few recent advances have showcased the potential for achieving remarkable performance in continuous speech decoding with extensive intracranial ECoG data collected through specific clinical trial protocols, combined with deep neural network models. Moses et al. \cite{moses2021neuroprosthesis} decoded spoken words with the word error rate (WER) at 25.6\% in a close vocabulary set (50 words in total) using 22 hours of recorded ECoG signals. Metzger et al. \cite{metzger2023high} synthesized natural speech with the phoneme error rate (PER) at 25\% using 17 days of recorded intracranial signals. Both studies used recurrent neural networks and language models. 

Conversely, small-scale datasets with 30 minutes of neural recordings, which is more feasible with general clinical settings, would only afford less complex DNN models with limited power of representation learning. Previous studies using these kinds of datasets have explored either classifying discrete phoneme classes \cite{mugler2014direct} or reconstructing the speech sound of isolated words \cite{pasley2012} using linear models. Deep neural network models, such as MLP, LSTM or transformers, have also been employed to reconstruct speech sounds of single digits \cite{akbari2019towards} and isolated words \cite{komeiji2022transformer} from intracranial signals. However, the inherent constraints of limited data make it infeasible to directly train end-to-end DNN models that directly map neural activity to full sentences of continuous natural speech with high fidelity and intelligibility.

In this work, we aim to bridge the gap between the limited amount of neural data and the complexity of representation learning required for natural speech reconstruction at the sentence level. To address this challenge, we introduce a novel transfer learning framework, called Neural2Speech. This framework leverages a pre-trained speech resynthesizer to improve the performance of sentence-level neural-driven speech reconstruction. Unlike previous neural-driven speech reconstruction works which trained the model on neural recording data from scratch, our proposed Neural2Speech approach consists of two training phases: a pre-training phase and a feature adaptor training phase. In the pre-training phase, we employ a speech autoencoder trained on readily accessible speech corpora. An important advantage of this pre-training phase is the ability to harness an enhanced speech representation learning model (speech encoder) with a substantial number of parameters. This allows us to overcome the constraints imposed by the availability of large neural recording data and obtain a strong speech resynthesizer (speech generator). In the feature adaptor training phase, in order to obtain a speech reconstruction model for the neural activity, we train a lightweight adaptor using a modest-sized neural recording dataset of 20 minutes. The primary function of this adaptor is to align the neural activity feature with the speech representations extracted from the pre-trained encoder, thereby facilitating the reconstruction of natural speech using the pre-trained speech generator.\footnote{Our code and the reconstructed speech waveforms are available at https://github.com/CCTN-BCI/Neural2Speech.}


The main contributions of our work are summarized as follows:
\begin{itemize}
    \item We propose a transfer learning framework for neural-driven speech reconstruction, called Neural2Speech.
    \item To overcome the limitation of neural record data, Neural2Speech adopts a pre-trained resynthesizer, trained on an easily accessible speech corpus, and a lightweight neural feature adapter, trained on modest-scale neural recording data.
    \item The effectiveness of Neural2Speech is validated through experiments using only 20 minutes of intracranial data, resulting in substantial performance improvements compared to existing baseline methods.
\end{itemize}
\end{sloppypar}

\begin{figure}[t]
\label{fig:overview}
\begin{minipage}[b]{1.0\linewidth}
  \centering
  \centerline{\includegraphics[width=8cm]{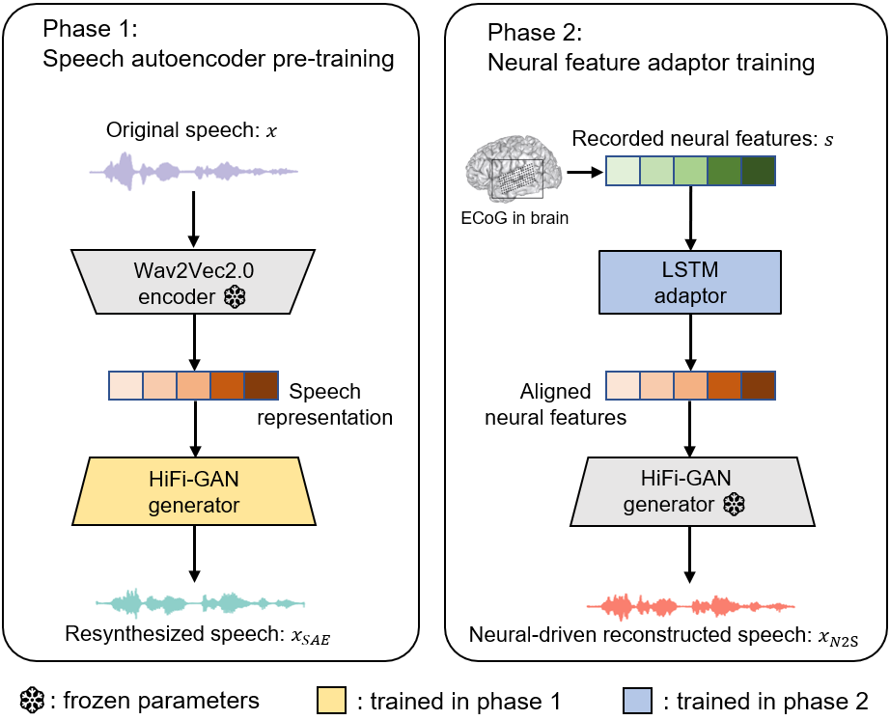}}
  \caption{Overview of the Neural2Speech framework. }
\end{minipage}
\end{figure}

\section{Methods}
\subsection{Overview of Neural2Speech}
According to the two aforementioned training phases, our proposed Neural2Speech framework (see \textbf{Fig.\ref{fig:overview}}) consisted of three modules: a deep speech encoder (Wav2Vec2.0 \cite{baevski2020wav2vec}), a deep speech generator (HiFi-GAN generator \cite{kong2020hifi}) and a neural feature adaptor. Notably, the Wav2Vec 2.0 encoder and the HiFi-GAN generator constituted the encoder-decoder components of the speech autoencoder. Wav2Vec2.0 was a representative pre-trained speech model, demonstrating superior performance in many speech tasks \cite{baevski2020wav2vec}. The HiFi-GAN was a generative adversarial network for efficient and high-fidelity speech synthesis \cite{kong2020hifi}, consisting of a generator and a discriminator. The discriminators were used to identify long-term dependencies, which is vital for natural speech. These modules enhanced the resemblance of generated speech to natural speech by focusing on long-term speech dependencies and human-sensitive frequency components. Finally, we used an LSTM module to align the recorded neural activity to the speech embedding features extracted from the Wav2Vec2.0 speech encoder. Previous work has shown that the speech embeddings in Wav2Vec 2.0 were strongly correlated to the speech responses in the human brain \cite{li2022dissecting}. Therefore, the LSTM module aligned the spatiotemporal neural features to the extracted speech features in Wav2Vec 2.0 encoder through nonlinear mapping, while maintaining a modest scale affordable by the limited amount of neural recording data. 

The entire model was successively trained in two phases: the speech autoencoder pre-training and the neural feature adaptor training. 
In the first phase, we froze the parameters of the encoder (Wav2Vec2.0 pre-trained on 960h LibriSpeech \cite{panayotov2015librispeech}) and trained the HiFi-GAN generator. In the neural feature adaptor training phase, we froze all the parameters of the pre-trained speech autoencoder and trained the LSTM alignment module.

\subsection{Speech autoencoder pre-training}
\begin{sloppypar}
In the first training phase, we pre-trained the speech autoencoder for speech reconstruction. The generator and the discriminators were adversarially trained \cite{kong2020hifi}. The loss function consisted of three parts: adversarial loss $L_{Adv}$, mel-spectrogram loss $L_{Mel}$ and feature matching loss $L_{FM}$:
\end{sloppypar}
\begin{equation}\label{eq:SAE_loss_total}
  L = L_{Adv}(G;D) + \lambda_{FM}L_{FM}(G,D) + \lambda_{Mel}L_{Mel}(G)
\end{equation}
We used the adversarial loss $L_{Adv}$ to train the HiFi-GAN discriminators to discriminate natural speech and resynthesized speech:
\begin{equation}\label{eq:SAE_loss_adv}
  L_{Adv}(D,G) = E_{(x)}[(D(x) - 1)^2 + (D(x_{SAE}))^2]
\end{equation}
where
\begin{equation}\label{eq:xSAE}
  x_{SAE} = G(E(x))
\end{equation}
$G$ and $D$ represented HiFi-GAN the generator and discriminator respectively, and $E$ represented the Wav2Vec2.0 encoder.

The mel-spectrogram loss $L_{Mel}$ was used to improve the training efficiency of the generator.
\begin{equation}\label{eq:SAE_loss_mel}
L_{Mel}(G) = E_{(x)}[\Vert \phi(x) - \phi(x_{SAE}) \Vert_1]
\end{equation}
where $\phi(x)$ was the mel-spectrogram of speech waveform $x$.

The feature matching loss $L_{FM}$ was a learned metric to measure the similarity between original speech and resynthesized speech:
\begin{equation}\label{eq:SAE_loss_fm}
L_{FM}(G,D) = E_{(x)}[\sum_{i=1}^{T} \frac{1}{N_i}\Vert D^i(x) - D^i(x_{SAE}) \Vert_1]
\end{equation}
where $T$ was the number of the layers in the discriminator, $D^i$ and $N_i$ were the features and the number of features in $i$-th layer of the discriminator.

In summary, the two additional terms of the loss function enhanced the resynthesized speech quality: 1) $L_{Mel}$ helped the generator in resynthesizing realistic waveforms. 
2) $L_{FM}$ addressed the properties of speech of different periods by using multiple small sub-discriminators.

\subsection{Neural feature adaptor training}
\justify
\begin{sloppypar}
In the second training phase, we trained the LSTM feature adaptor $F$ to align the ECoG recording signals to the pre-trained speech features. Alternative non-sequential adaptors, such as MLPs, exhibited notably inferior performance. This observation led us to infer the critical importance of temporal information within the neural activity for effective feature alignment. Therefore we focused on sequential structures for the adaptor. 

Within the training of $F$, we froze all the parameters of the pre-trained speech autoencoder. The loss function used in this phase consisted of two parts: the mel-spectrogram loss $L_{Mel}$ and the Laplacian loss $L_{Lap}$:
\begin{equation}\label{eq:LSTM_loss_diff}
  L = L_{Mel} + \lambda L_{Lap} 
\end{equation}
The mel-spectrogram loss $L_{Mel}$ was used to improve the acoustic fidelity and intelligibility of the final reconstructed speech from recorded neural activity: 
\begin{equation}\label{eq:LSTM_loss_lat}
  L_{Mel} = \Vert \phi(x) - \phi(x_{N2S}) \Vert _{1}
\end{equation}
where
\begin{equation}
x_{N2S} = G(F(s))
\end{equation}
$x$ and $x_{N2S}$ represented the raw speech and the neural-driven reconstructed speech, and $s$ represented the recorded neural activity.

Moreover, in order to improve the phonetic intelligibility, we applied the Laplace operator as a convolution kernel to the mel-spectrogram, which was a simplified representation of the convolution on the spectrogram, while convolution on spectrogram was proven to be effective on speech denoising and enhancement \cite{xuhong2021speech}:  
\begin{equation}\label{eq:LSTM_loss_mel}
  L_{Lap} = \Vert \Delta(\phi(x)) - \Delta(\phi(x_{N2S})) \Vert_{2}
\end{equation}
where $\Delta$ represented the Laplacian operator on the mel-spectrogram.

In summary, the two terms of the loss function contributed to the training: 1) Similar to the pre-training of the speech autoencoder, $L_{Mel}$ helped to improve the acoustic fidelity of the reconstructed speech. 2) $L_{Lap}$ helped to improve the phonetic intelligibility of reconstructed speech efficiently.

\end{sloppypar}

\section{Experiments and results} 
\subsection{Experimental setup}
\subsubsection{Data}
\label{sec:data}
{\bf Neural recording data.} We recorded the neural activity of nine right-handed participants using high-density ECoG grids while they passively listened to 
599 sentences of natural speech, a 20-minute subset from the TIMIT corpus \cite{TIMIT} during their surgical treatment of epilepsy.\footnote{The experiment was approved by the Institutional Review Board at UCSF. Participants gave full consent approval before the experiments.}
The ECoG grids mainly covered the speech auditory cortex. The pre-processing of recorded neural activity consisted of three steps: filtering, speech-sensitive electrode selection, and down-sampling. Firstly, we filtered the high-gamma band signals (70-150Hz) from the recorded neural activity because of its high correlations to speech \cite{pei2011spatiotemporal}. Secondly, we used a paired sample $t$-test to examine whether the average post-onset response (400 ms to 600 ms) was higher than the average pre-onset response (-200 ms to 0 ms, $p < 0.01$, $t$-test, one-sided, Bonferroni corrected), in order to select speech-responsive electrodes. Finally, we down-sampled the filtered neural activity to the same as the speech embedding sequence from Wav2Vec2.0 (49Hz) for feature alignment.

\noindent {\bf Speech corpus.} We used the public 960 hours of LibriSpeech corpus \cite{panayotov2015librispeech} for the pre-training of the speech autoencoder. 
\begin{center}
\begin{table}[]\label{tab:SAE}
\caption{Fidelity and intelligibility evaluations of the pre-trained speech autoencoder on TIMIT. (Ground truth corresponds to the metrics evaluated on the original raw speech.)} 
\begin{tabularx}{0.48\textwidth}{ccc}
\hline
\makebox[0.14\textwidth][c]{} & 
\makebox[0.10\textwidth][c]{Ground truth} & 
\makebox[0.16\textwidth][c]{Resynthesized speech}       \\ \hline
Mel-spectral MSE  & 0               & 0.200$\pm$0.043   \\ \hline
MOS           & 4.287$\pm$0.321 & 3.543$\pm$0.409    \\ \hline
ESTOI         & 1               & 0.825$\pm$0.045   \\ \hline
\end{tabularx}
\end{table}    
\end{center}
\subsubsection{Model and training}
{\bf Model.} Our Neural2Speech model mainly consisted of a pre-trained Wav2Vec2.0 encoder, a HiFi-GAN generator, and an LSTM adaptor. The Wav2Vec2.0 consisted of 7 1D convolutional layers which downsampled 16kHz natural speech to 49Hz, extracting 768-dimensional local features, and 12 transformer-encoder blocks, extracting contextual feature representations from unlabeled speech data. The HiFi-GAN generator incorporated 7 transposed convolutional layers for upsampling and multi-receptive field (MRF) fusion modules. A single MRF module added the output features from multiple ResBlocks. The discriminator employed both multi-scale and multi-period discriminators. The upsample rates of each ResBlock in the MRFs were set as [2, 2, 2, 2, 2, 2, 5] respectively. The upsample kernel sizes of each ResBlock in the MRFs were set as [2, 2, 3, 3, 3, 3, 10]. The periods of the multi-period discriminator were set as [2, 3, 5, 7, 11]. The neural feature adaptor consisted of 2 unidirectional LSTM layers. The adaptor took in the preprocessed neural signals as the input. The dimensions of the hidden space and the output were 300 and 768 respectively.

\noindent {\bf Training.} Neural2Speech was trained in two phases. In the first pre-training phase, the parameters of the HiFi-GAN generator were trained using Adam optimizer \cite{kingma2014adam} with a fixed learning rate as $1 \times 10^{-5}$, $\beta_1$ = 0.9, $\beta_2$ = 0.999. We set $\lambda_{FM} = 2$ and $\lambda_{Mel} = 50$ in Eq. (\ref{eq:SAE_loss_total}). We trained this autoencoder on 960h LibriSpeech \cite{panayotov2015librispeech} corpus for 400 epochs, which took 12 days on 6 NVidia Geforce RTX3090 GPUs. The parameters of the LSTM were trained using a stochastic gradient descent (SGD) optimizer with a fixed learning rate as $3\times10^{-3}$ and a momentum as 0.8. We set $\lambda = 0.1$ in Eq. (\ref{eq:LSTM_loss_diff}). The normalization coefficient $\lambda$ was selected using an ablation test. For each participant, we trained a personalized feature adaptor. Each single feature adaptor was trained for 500 epochs on one NVidia Geforce RTX3090 GPU, which took 12 hours. The train, validation, and test sets took up 70\%, 20\%, and 10\% of the total trials (599 sentences) respectively.

\subsubsection{Evaluation metrics}
We evaluated the quality of the resynthesized speech from the pre-trained speech autoencoder using fidelity (Mean Square Error (MSE) of the mel-spectrogram) and intelligibility indicators (Mean opinion score (MOS) \cite{salza1996mos} and extended short-time objective intelligibility (ESTOI) \cite{jensen2016algorithm}) respectively on the TIMIT test set.
MOS score is a subjective metric to evaluate the intelligibility of speech. We asked 20 volunteers who have had at least 10 years of English learning experience to rate the quality of synthesized speech from the pre-trained speech autoencoder on a numerical scale: 1 (bad), 2 (poor), 3 (fair), 4 (good), and 5 (excellent). 
ESTOI score is an objective indicator to assess the intelligibility of speech synthesis tech ranges from 0 (worst) to 1 (best). It measures the distortion in spectrotemporal modulation patterns of speech, sensitive to spectral and temporal pattern inconsistencies.

We evaluated the performance of neural-driven reconstructed speech using the MSE of mel-spectrogram, ESTOI score, and phoneme error rate (PER). The mel-spectrogram MSE and ESTOI were similar to the previous training phase. The PER measured the percentage of phonemes incorrectly or not recognized at all by the phoneme recognizer. It was calculated by comparing the edit distance between the reference transcription and the phoneme recognizer output.

\begin{table}[t]\label{tab:rec}
\caption{The performance of neural-driven speech reconstruction using different models.}
\centering
\begin{tabularx}{0.48\textwidth}{Xcc}
\hline
\makebox[0.16\textwidth][c]{} 
& \makebox[0.14\textwidth][c]{Mel-spectral MSE}
& \makebox[0.1\textwidth][c]{ESTOI} \\ \hline
 \centering
MLP regression \cite{bellier2023music}          & 29.186±7.426 & 0.067±0.008 \\ \hline
 \centering
LSTM-CNN regression & 0.104±0.002 & 0.063±0.007 \\ 
\hline
 \centering
Our Neural2Speech & 0.067±0.018 & \textbf{0.371±0.013}   \\

\ \ \ \ \ \ \ \ \ \ \ \ w/o $L_{Lap}$& \textbf{0.017±0.005} & 0.327±0.011 \\ 
\ \ \ \ \ \ \ \ \ \ \ \ w/o $L_{Mel}$& 0.150±0.039   & 0.308±0.013   \\ \hline


\end{tabularx}
\end{table}

\begin{figure}[t]\label{fig:rec}
\begin{minipage}[b]{1.0\linewidth}
  \centering
  \centerline{\includegraphics[width=6.5cm]{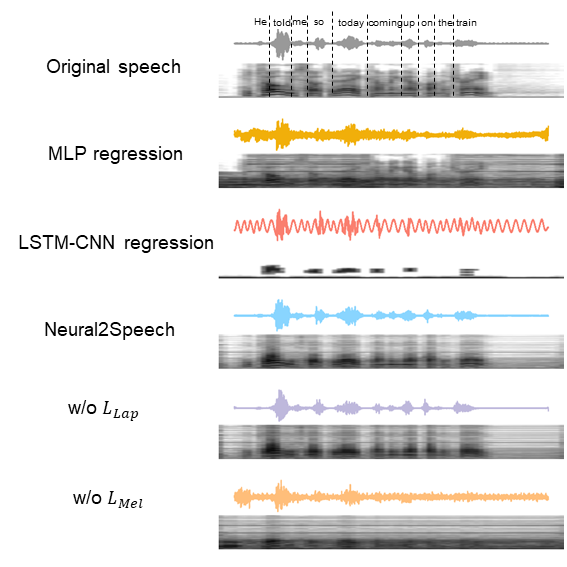}}
  \caption{Visualization of the speech waveform and the mel-spectrogram of the raw speech and speech reconstructed from ECoG recordings using different methods for a sample sentence.}
\end{minipage}
\end{figure}

\subsection{Results and Discussions}
\subsubsection{Performance of the pretrained autoencoder}
We first evaluated the performance of the pre-trained speech autoencoder on the TIMIT test set, which consisted of 1680 sentences. We compared the quality of the resynthesized speech to the ground truth. The performance of our pre-trained speech autoencoder was shown in \textbf{Table 1}. Using both objective and subjective metrics, we found that the resynthesized speech showed satisfactory fidelity and intelligibility close to the ground truth, with mean MSE = 0.20, MOS = 3.54 and ESTOI = 0.825. 


\subsubsection{Performance of Neural2Speech}
Next, we evaluated the final neural-driven speech reconstruction from the complete Neural2Speech framework with actual neural recording data. As shown in \textbf{Table 2}, the Neural2Speech reconstructed speech showed high fidelity and high intelligibility, with mean MSE = 0.067, and mean ESTOI = 0.371, averaged across all test data and across 9 human participants. In particular, we achieved a mean ESTOI of 0.395 ± 0.014 for the best participant. 

To highlight the importance of the pre-trained speech autoencoder in our framework, we compared our model with two baseline models (\textbf{Table 2 and Fig. 2}). We trained an MLP-based neural-to-speech decoder from scratch using only the 20-min ECoG data (the MLP regression in \textbf{Table 2}), similar to the methods used in a recent study \cite{bellier2023music}. We also trained a sequence-to-sequence LSTM-CNN regression model from scratch using only the 20-min ECoG data, a framework similar to \cite{makin2020machine}. Neither of the two baselines achieved a reasonable MSE or ESTOI value. Therefore, representation learning through large-scale pre-training is vital to neural-driven speech reconstruction, particularly under the constraint of limited neural recording data. 

Furthermore, we performed ablation analyses of the Neural2Speech training. \textbf{Table 2} presents the results using only $L=L_{Mel}$ or $L=L_{Lap}$. Specifically, $L_{Mel}$ contributed most to the acoustic fidelity of the reconstructed speech, quantified by the Mel-spectral MSE; both $L_{Mel}$ and $L_{Lap}$ enhanced the phonetic intelligibility (ESTOI) of the reconstructed speech.


\begin{figure}[]\label{fig:per}
\begin{minipage}[b]{1.0\linewidth}
  \centering
  \centerline{\includegraphics[width=6.5cm]{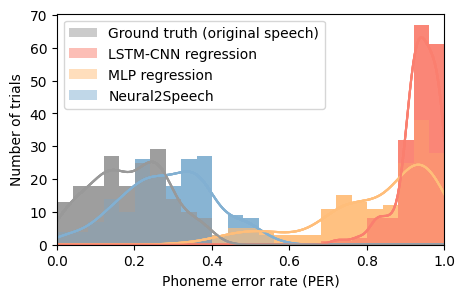}}
  \caption{Transcription PER for individual trials on different reconstruction methods and their corresponding Kernel Density Estimation (KDE) curves.}
\end{minipage}
\end{figure}

To further quantitatively evaluate the intelligibility of reconstructed speech, we trained an automatic phoneme recognizer that transcribes speech waveform into phoneme sequences, using the TIMIT set. This recognizer was fined-tuned from a pre-trained Wav2Vec2.0 model. \textbf{Fig.3} shows the histograms of the PER distribution evaluated on 180 test trials (20 trials for each participant). 
The average PER of the neural-driven reconstructed speech was 0.286 ± 0.110 (best performance across participants: 0.261 ± 0.103), which was close to the performance evaluated on the ground truth (the raw speech PER = 0.189 ± 0.082). None of the other baseline models achieved PER below 0.5. These results further validate the intelligibility of the neural-driven speech reconstruction. 

\section{Conclusion}
In this paper, we propose a transfer learning framework for reconstructing sentence-level intelligible input speech from limited intracranial neural recordings, a more complex task compared to prior works focusing on phonemes, single digits, or words. Our approach leverages pre-trained speech autoencoder and neural feature alignment modules to address the challenges of complexity in speech representation learning and data scarcity. 


This framework allows us to achieve sentence-level speech decoding with high fidelity and intelligibility with only 20 minutes of neural data, significantly less than previous approaches requiring hours or days of recordings. Notably, our Neural2Speech framework outperforms previous methods, particularly when compared to training from scratch. The potential applications of this framework in BCI domain hold promise for further advancements in this field.

\newpage
\bibliographystyle{IEEEbib}
\bibliography{refs}


\end{document}